\documentclass[aps]{revtex4}
\usepackage{amssymb}
\usepackage{amsmath}
\usepackage{graphicx}
\newtheorem{theorem}{Theorem}[section]

\newcommand{\qed}{\hfill \hbox{\rule[-2pt]{3pt}{6pt}}}

\def\cA{{\mathcal A}}

\def\>{\rangle}
\def\<{\langle}

\def\0{{\mathbf 0}}

\def\({\left(}
\def\){\right)}
\def\[{\left[}
\def\]{\right]}

\begin{document}
\openup 1.3 \jot

\title[Noiseless Subsystem and Decoherence Free Subspace]
{Recursive Encoding and Decoding of Noiseless Subsystem \\
and Decoherence Free Subspace}

\author{Chi-Kwong Li}
\address{Department of Mathematics, College of William \& Mary,
Williamsburg, VA 23187-8795, USA. (Year 2011: Department of Mathematics, 
Hong Kong University of Science \& Technology, Hong Kong.)}
\email{ckli@math.wm.edu}

\author{Mikio Nakahara}
\address{Research Center for Quantum Computing,
Interdisciplinary Graduate School of Science and Engineering,
and Department of Physics, Kinki University,
Kinki University, 3-4-1 Kowakae, Higashi-Osaka, 577-8502, Japan.}
\email{nakahara@math.kindai.ac.jp}

\author{Yiu-Tung Poon}
\address{Department of Mathematics, Iowa State University,
Ames, IA 50051, USA.}
\email{ytpoon@iastate.edu}

\author
{Nung-Sing Sze}
\address{Department of Applied Mathematics, The Hong Kong Polytechnic 
University, Hung Hom, Hong Kong.}
\email{raymond.sze@inet.polyu.edu.hk}

\author{Hiroyuki Tomita}
\address{Research Center for Quantum Computing,
Interdisciplinary Graduate School of Science and Engineering,
Kinki University, 3-4-1 Kowakae, Higashi-Osaka, 577-8502, Japan.}
\email{tomita@alice.math.kindai.ac.jp}






\begin{abstract}
When the environmental disturbace to a quantum system has a wavelength
much larger than the system size, all qubits localized within a small area
are under action of the same error
operators. Noiseless subsystem and decoherence free subspace are known
to correct such collective errors. We construct simple quantum circuits, 
which implement these collective error correction codes, for a small number $n$
of physical qubits. A single logical qubit is encoded with $n=3$
and $n=4$, while two logical qubits are encoded with $n=5$. 
The recursive relations among the subspaces employed in noiseless subsystem
and decoherence free subspace play essential r\^oles in our implementation.
The recursive relations also show that the number of gates required to encode
$m$ logical qubits increases linearly in $m$.
\end{abstract}

\maketitle

\section{Introduction}
\setcounter{equation}{0}

A quantum system is vulnerable to external noise. In quantum information 
processing and quantum computation, the system must be protected
from the environmental noise one way or another
to protect information stored in the quantum registers.
The majority of quantum systems employed for these purposes
is microscopic in size, typically on the order of a few microns.
In contract, the environmental noise, such as electromagnetic wave,
has the wavelength on the order of a few centimeters or more. 
Therefore, it is natural to assume all the qubits in the register
suffer from the same error operator. We call such error 
the collective error in the following. Suppose $n$-qubit quantum states
$\rho$ are represented as $N\times N$ density matrices with $N = 2^n$,
and a quantum channel is realized as a completely positive linear map
$\Phi$ with an operator sum representation
\begin{equation} \label{channel}
\Phi(\rho) = \sum_{j=1}^r E_j \rho E_j^*
\end{equation}
for the error operators $E_1, \dots, E_r$; see \cite{NO,NC}. 
Then the error operators of our channel can be expressed as multiples of 
operator of the form $W^{\otimes n} \in {\bf{2}}^{\otimes n}$, where 
${\bf{2}}$ is the two-dimensional (fundamental) 
irreducible representation of SU(2).

Decoherence free subspace \cite{dfs1,dfs2,dfs3,dfs4}
and noiseless subsystem \cite{ns1,ns2,ns3,Kempe} 
are two standard
methods to correct collective errors; see \cite{Kempe,Kribs}.
It is not hard to explain the scheme using the operator sum representation
of the quantum channel (\ref{channel}) as follows. Suppose the 
finite dimensional $C^*$-algebra $\cA$
generated by the error operators admits the unique decomposition into
irreducible representations up to unitary
equivalence (similarity) as
$$\bigoplus_{j} (I_{r_j} \otimes M_{n_j}) \qquad \hbox{ with } \ \sum_j r_j n_j = N,$$
where $n_j$ is the dimension of the irreducible representation while
$r_j$ its multiplicity.
Then every error operator $E_i$ in (\ref{channel}) has the form
$$\bigoplus_{j} (I_{r_j} \otimes B_j) \qquad \hbox{ with } \ B_j \in M_{n_j}.$$
For every index $j$, if we regard 
$$M_N = (I_{r_j} \otimes M_{n_j}) \oplus M_{q}, \quad q = N - r_jn_j,$$
and if apply the channel to a quantum state $\rho = (\hat \rho \otimes \sigma) \oplus O_q$ with $\hat \rho \in M_{r_j}$ and $\sigma \in M_{n_j}$,
according to this decomposition, then 
$$\Phi(\rho) = (\hat \rho \otimes \sigma_E) \oplus O_q$$
because of the special form of the error operators in this decomposition.
Here $O_q$ is a null matrix of order $q$.
Thus, the state $\hat \rho \in M_{r_j}$ encoded as above will not be affected by
the errors (noise) and can be easily recovered.
This gives rise to a noiseless subsystem. 
The situation is particularly pleasant if $n_j = 1$, i.e., we use the one dimensional
irreducible representation of $\cA_n$, so that
$$\Phi(\hat \rho \oplus O_q) = \hat \rho \oplus O_q.$$
In such a case, we get a decoherence free subspace.

We are interested in an efficient construction, 
which leads to simple implementation, 
of decoherence free subspaces and noiseless subsystems for the channels with
common error on each qubit in the register.  By the discussion in the preceding
paragraph, construction of decoherence free subspace employs 
one-dimensional irreducible representations of the algebra $\cA_n$ 
generated by 
${\bf{2}}^{\otimes n}$ for encoding while the latter encodes logical qubits by 
making use of the multiplicity of some irreducible representations. 

It is the purpose of this paper to investigate the implementation of these
ideas in terms of the quantum circuits. We consider decoherence free
subspace with $n=4$, which implements a single logical qubit and
noiseless subsystem with $n=3$ and $5$, which encodes a single
logical qubit and two logical qubits, respectively.
Viola {\it et al} \cite{Viola} worked out the circuit implementation of $n=3$ noiseless 
subsystem and demonstrated its validity by using ion trap quantum computer.
No further works have been conducted for $n \geq 4$ to date to our knowledge.
Our implementation, starting with $n=3$ noiseless subsystem, is recursive
so that $n=4$ decoherence subsystem and $n=5$ noiseless subsystem are
implemented with the quantum circuit for $n=3$. Moreover, our circuit
for $n=3$ is simpler than that obtained by Yang and Gea-Banacloche \cite{ns3}
and by Viola {\it et al} \cite{Viola}.

We construct a quantum circuit for $n=3$ noiseless subsystem in 
the next section. We analyze $n=4$ decoherence free subspace
and $n=5$ noiseless subsystem in Sections III and IV
by making use of the result of Section II.
Our analysis is concrete and
encoding basis vectors and quantum circuits are explicitly
constructed. The last section is devoted to summary and discussion.

We will use the known fact (see \cite{Kempe}) 
that the algebra $\cA_n$ generated by ${\bf{2}}^{\otimes n}$ 
has the unique decomposition
$$\bigoplus_{0 \le j \le n/2} (I_{r_j} \otimes M_{n_j})$$
with $(r_0, n_0) = (1, n+1)$ and $(r_j, n_j) = \left({n\choose j}-{n\choose j-1}, n+1-2j\right)$ 
for $0 < j \leq n/2$. Also, we will employ the Lie theoretic notation
and regard a qubit belonging to the fundamental 
representation $\bf{2}$ 
of SU(2) while the product operator
$W^{\otimes n}$ acts as a reducible representation of SU$(2)^{\otimes n}$, denoted by
${\bf 2}^{\otimes n}$.

\section{3-qubit Noiseless Subsystem}

Let us consider a 3-qubit system and see how it can be used
to encode a logical qubit which is robust against any noise
of the form $W^{\otimes 3}$, where $W$ is an arbitrary element
of the fundamental representation $\bf{2}$. 
To this end, we first consider the algebra $\mathcal{A}_3$
of ${\bf{2}}^{\otimes 3}$.
$\mathcal{A}_3$ 
is decomposed into the sum of irreducible representations as
$$
{\bf{2}}^{\otimes 3} ={ \bf{4}} \oplus (I_2 \otimes {\bf{2}}),
$$
where $I_n$ is the unit matrix of dimension $n$.
Corresponding to this decomposition, any unitary matrix 
$V \in {\bf{2}}^{\otimes 3}$
can be decomposed as
$$
V= V_4 \oplus (I_2 \otimes V_2)
$$
under a proper choice of basis vectors. Here $V_4$ belongs to $\bf{4}$
and $V_2$ to $\bf{2}$ of SU(2). It should be noted that $I_2$ is immune to
any collective noise of the form $W^{\otimes 3}, W \in \bf{2}$ and the 
corresponding vector space form the noiseless subsystem.

The success of our schemes depends on a judicious choice of orthonormal
basis for the decomposition of the algebra $\cA$ generated by 
${\bf{2}}^{\otimes 3}$. 
To this end, let $\{|e_{4,1}\rangle, |e_{4,2}\rangle, |e_{4,3}\rangle,
 |e_{4,4}\rangle |\}$ 
be a basis of $\bf{4}$, $\{|e_{a1} \rangle, |e_{a2} \rangle \}$ and
$\{|e_{b1} \rangle, |e_{b2} \rangle \}$ be the bases of the two ${\bf 2}$ defined as follows
\footnote{Let $|s, m \>$ denote the eigenvector of $S^2$ and $S_z$ with 
eigenvalues $s(s+1)$ and $m$, respectively. Then
$|e_{4,k}\rangle$ corresponds to $|\frac{3}{2},
\frac{5}{2}-k \rangle$ and $|e_{a(b)k}\rangle$ to $|\frac{1}{2},
\frac{3}{2}-k \rangle$.
Explicitly, they are
$|e_{4,1} \rangle = |\frac{3}{2}, \frac{3}{2} \rangle$,
$|e_{4,2} \rangle = |\frac{3}{2}, \frac{1}{2} \rangle$,
$|e_{4,3} \rangle = |\frac{3}{2}, -\frac{1}{2} \rangle$,
$|e_{4,4} \rangle =|\frac{3}{2}, -\frac{3}{2} \rangle$,
$|e_{a1} \rangle =|\frac{1}{2}, \frac{1}{2} \rangle$,
$|e_{a2} \rangle =|\frac{1}{2}, -\frac{1}{2} \rangle$,
$|e_{b1} \rangle =|\frac{1}{2}, \frac{1}{2} \rangle$,
$|e_{b2} \rangle =|\frac{1}{2}, -\frac{1}{2}\rangle$.}.
\begin{equation}
\left\{
\begin{array}{l}
\displaystyle |e_{4,1} \rangle = |000\rangle,\\
\displaystyle |e_{4,2} \rangle = \frac{1}{\sqrt{3}}(|100\rangle+|010\rangle+|001\rangle),\\
\displaystyle |e_{4,3} \rangle =  \frac{1}{\sqrt{3}}(|011\rangle+|101\rangle+|110\rangle),\\
\displaystyle |e_{4,4} \rangle = |111\rangle,
\end{array} \right.
\end{equation}
\begin{equation}
\left\{
\begin{array}{l}
\displaystyle |e_{a1} \rangle = \frac{1}{\sqrt{2}} (|100 \rangle-|010\rangle),\\
\displaystyle |e_{a2} \rangle =-\frac{1}{\sqrt{2}} (|011 \rangle-|101 \rangle),
\end{array}
\right.
\end{equation}
\begin{equation}
\left\{
\begin{array}{l}
\displaystyle |e_{b1} \rangle = \frac{1}{\sqrt{6}}(|100\rangle+|010\rangle-2|001\rangle),\\
\displaystyle |e_{b2} \rangle = - \frac{1}{\sqrt{6}}(|011\rangle+|101\rangle-2|110\rangle).
\end{array} \right.
\end{equation}
We implement a noiseless subsystem from
two $\bf{2}$ representations.

Suppose $U_E^{(3)}$ 
is an encoding matrix which generates the above basis vectors
from the binary basis vectors $|i_1i_2i_3 \rangle,\ (i_k \in \{0, 1\})$.
We choose $U_E^{(3)}$ to have columns
$$
(|e_{a1}\rangle, |e_{b1} \rangle, |e_{4,2} \rangle, |e_{4,1} \rangle,
|e_{a2}\rangle, |e_{b2} \rangle, |e_{4,3} \rangle, |e_{4,4} \rangle)
$$
in this order. 

\begin{theorem}
Let $\alpha, \beta, \gamma$ be any real numbers and let
$$
X_{\alpha} = (e^{i \alpha \sigma_x})^{\otimes 3},
Y_{\beta} = (e^{i \beta \sigma_y})^{\otimes 3},
Z_{\gamma} = (e^{i \gamma \sigma_z})^{\otimes 3},
$$
where $\sigma_k$'s are the Pauli matrices.
Consider a quantum channel $\Phi:M_8 \to M_8$ given by
$$
\Phi(\rho) = p_0 \rho + p_1 X_{\alpha} \rho X_{\alpha}^{\dagger}
+p_2 Y_{\beta} \rho Y_{\beta}^{\dagger}
+p_3 Z_{\gamma} \rho Z_{\gamma}^{\dagger}
$$
for some $p_i \in \mathbb{R}$ such that $\sum_{i=0}^3 p_i \leq 1$.
Then for any data state $\hat{\rho} \in M_2$, $U_E^{(3)}$ and $\Phi$ satisfy
the identity
\begin{equation}\label{eq:3}
U_E^{(3)\dagger} \Phi\left(U_E^{(3)}(\rho_a \otimes 
|0 \rangle \langle 0| \otimes \hat{\rho})
U_E^{(3)\dagger}\right) U_E^{(3)}
= \left(\sum_{j=0}^3 p_j U_j \rho_a  U_j^{\dagger} \right)
\otimes|0 \rangle \langle 0|\otimes \hat{\rho},
\end{equation}
that is, the initial data state is recovered in the output state with no 
entanglement with the ancilla qubits. Here
$\rho_a$ is an initial single qubit ancilla state and
$$
U_0 = I_2,\ U_1= e^{i \alpha \sigma_x},\ U_2= e^{i \beta \sigma_y},\ U_3
= e^{i \gamma \sigma_z}.
$$
\end{theorem}
\medskip
Proof: We show that the $\bf{2}\oplus \bf{2}$ irreducible representations form
a noiseless subsystem by explicit evaluation.
Let $\{|e_{a1}\rangle, |e_{a2}\rangle\}$ spans the logical 
$|0 \rangle_L$ state,
while $\{|e_{b1} \rangle , |e_{b2} \rangle\}$ spans the logical 
$|1 \rangle_L$ state.
We show that noise operators $X_{\alpha}, Y_{\beta}$ and $Z_{\gamma}$
leave each subspace invariant.

Let $P_a = \sum_{i=1}^2 |e_{ai} \rangle \langle e_{ai}|$ and 
$P_b = \sum_{i=1}^2 |e_{bi} \rangle \langle e_{bi}|$. Then it is easy to
show
$$
X_{\alpha} P_k X_{\alpha}^{\dagger} =Y_{\beta} P_k Y_{\beta}^{\dagger} =
Z_{\gamma} P_k Z_{\gamma}^{\dagger} = P_k\ (k=a, b).
$$
It should be noted that, although
the whole four-dimensional subsystem is invariant
under $X_{\alpha}, Y_{\beta}$ and $Z_{\gamma}$, we cannot use this 
subsystem to encode two-qubit state since each vector is not invariant 
under the action of the error operators. 

Now it is easy to prove the identity. We use a pure state notation to simplify
the expressions.
The general case with mixed initial states $\rho_a$ and $\hat{\rho}$
is obtained by simply mixing the pure state results using linearity.
Let $|\hat{\psi} \rangle = a|0\rangle + b|1 \rangle$ be
a data qubit state to be encoded and $|v \rangle = v_0|0 \rangle + v_1
|1 \rangle$
be the initial state of the first ancilla qubit, while that of
the second qubit is set to $|0 \rangle$.
Under the action of $U_E^{(3)}$, along with a two qubit
state $|v\rangle |0\rangle$, $|\hat{\psi} \rangle$ is encoded as
$$
|\Psi \rangle =
U_E^{(3)}|v \rangle |0 \rangle |\hat{\psi} \rangle = v_0
(a|e_{a1} \rangle +
 b |e_{b1}\rangle) + v_1 (a|e_{a2} \rangle + b |e_{b2}\rangle).
$$
Let us consider a noise operator $X_{\alpha}$ first. Its action on 
$|\Psi \rangle$ yields
\begin{eqnarray*}
|\Psi_X \rangle &=& 
X_{\alpha} |\Psi \rangle\\
& =& (v_0 \cos \alpha + i v_1 \sin \alpha)(a|e_{a1} \rangle + b|e_{b1}\rangle)
+ (v_1 \cos \alpha + i v_0 \sin \alpha) 
(a|e_{a2} \rangle + b|e_{b2}\rangle).
\end{eqnarray*}
The action of the recovery operator 
$U_E^{(3)\dagger}$ recovers the initial state, except
for the first qubit, as
$$
U_E^{(3)\dagger} 
|\Psi_X\rangle = \left( e^{i \alpha \sigma_x}|v \rangle \right)
 |0 \rangle |\hat{\psi} \rangle,
$$
which shows that data qubit state is immune to $X_{\alpha}$. It is shown
similarly that the data qubit is immune to other error operators either.
Since each error is in action with the probability $p_i$, we have proved 
the identity (\ref{eq:3}). \qed

A remark is in order. In contrast with an ordinary QECC,
the scheme corrects multiple action of the error operators.
It was shown in the theorem that the top-most qubit can be any
superposition state or mixed state initially and its output state is 
another superposition/mixed state
under an action of a single collective error operator in $X_{\alpha},
Y_{\beta}$ and $Z_{\gamma}$. It should be noted that the error channel
leaves the encoded word unchanged. Namely, given any initial ancilla state
$\rho_a$, there exists an ancilla state $\rho'_a$ such that
$$
\Phi\left(U_E^{(3)}(\rho_a \otimes |0\>\<0| \otimes \hat{\rho})U_E^{(3)\dagger}\right)
= U_E^{(3)}(\rho'_a \otimes  |0\>\<0| \otimes \hat{\rho}) U_E^{(3)\dagger}.
$$
Then the error correction may be repeated as many times as required. 
This implies that it corrects any error operator of the form
$W^{\otimes 3}$, where $W \in {\bf{2}}$. This is because any
element $W\in \bf{2}$ of SU(2) is decomposed into a product
$$
W= e^{i \theta_1 \sigma_x} e^{i \theta_2 \sigma_y} e^{i \theta_3 \sigma_x}.
$$
It should be clear that
$W^{\otimes 3}$ is expressed as a product $X_{\theta_1} Y_{\theta_2}
X_{\theta_3}$, each factor of which leaves the noiseless subsystem
invariant.

\begin{figure}
\includegraphics[width=7cm]{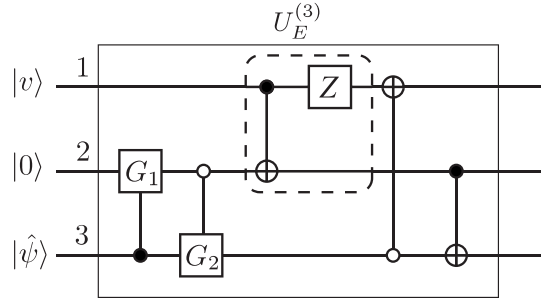}
\caption{Encoding circuit $U_E^{(3)}$ of the noiseless subsystem for a 3-qubit
system. The filled (empty) circle attached to the control qubit denotes that
the gate acts on the target qubit when the control qubit is set to $|1\>$ 
($|0\>$), respectively, and otherwise it is left alone.
It encodes a single qubit state $|\hat{\psi}\rangle$.
See text for redefinition of basis vectors to simplify the circuit.
The part surrounded by a broken line can be omitted if the initial state
of the top-most qubit is $|0 \rangle$, which makes the circuit even simpler.
The recovery operation is given by $U_E^{(3)\dagger}$. This circuit is 
employed as a module
in the implementation of the noiseless subsystem and the decoherence
free subspace for larger $n$.}\label{fig:3}
\end{figure}

One of the simplest quantum circuits which implement the encoding matrix 
$U_E^{(3)}$ is 
obtained by simple redefinitions of the basis vectors;
\begin{eqnarray*}
|e_{a1} \rangle &=& \frac{1}{\sqrt{2}}(|100 \rangle - |001 \rangle),\\
|e_{b1} \rangle &=& \frac{1}{\sqrt{6}}(|100 \rangle + |001 \rangle
-2 |010 \rangle),\\
|e_{4,2} \rangle & =& |111\rangle,\\
|e_{4,1} \rangle &=& \frac{1}{\sqrt{3}}(|100 \rangle + |001\rangle
+|010 \rangle),\\
|e_{a2} \rangle &=& -(\sigma_x)^{\otimes 3}|e_{a1} \rangle,
\ |e_{b2} \rangle = -(\sigma_x)^{\otimes 3}|e_{b1} \rangle,\\
|e_{4,3} \rangle &=& -(\sigma_x)^{\otimes 3}|e_{4,2} \rangle,
\ |e_{4,4} \rangle = -(\sigma_x)^{\otimes 3}|e_{4,1} \rangle.
\end{eqnarray*}
A permutation of the basis vectors takes much simpler form with the 
redefined basis and the quantum circuit is found by inspection. 
Figure~\ref{fig:3} shows an example of the encoding circuit, in which
$G_1$ and $G_2$ stand for
$$
G_1 = \frac{1}{\sqrt{3}} \left( \begin{array}{cc}
1&\sqrt{2}\\
-\sqrt{2}&1
\end{array}\right),\quad G_2 = \frac{1}{\sqrt{2}} \left( \begin{array}{cc}
1&1\\
-1&1
\end{array}\right).
$$
Note that our circuit
is simpler than that found in \cite{ns3} and \cite{Viola} regarding
the number of gates. 

\section{4-qubit Decohrence Free Subspace}

We design the 4-qubit decoherence free subspace, which is robust against
collective noise of the form $W^{\otimes 4} \ (W \in {\bf 2})$, by taking
advantage of the noiseless subsystem analyzed in the previous section.

A 4-qubit system is used to encode a logical qubit which is robust against 
any collective noise.
The algebra $\mathcal{A}_4$ obtained from $\bf{2}^{\otimes 4}$
is decomposed into the sum of irreducible representations; 
$$
{\bf{2}}^{\otimes 4} ={ \bf{5}} \oplus (I_3 \otimes {\bf{3}})
\oplus (I_2 \otimes {\bf{1}}).
$$
Corresponding to this decomposition, any unitary matrix 
$V \in {\bf{2}}^{\otimes 4}$ can be decomposed as
$$
V = V_5 \oplus (I_3 \otimes V_3) \oplus (I_2 \otimes V_1)
$$
under a proper choice of basis vectors. Here $V_k$ belongs to the
irreducible representation $\bf{k}$,
$k=1, 3, 5$ of SU(2). It should be noted that the singlet irreducible
representation is immune to any operator $V=W^{\otimes 4}, W \in {\bf{2}}$
and two of them form a single logical qubit which is immune to any noise
of the form $V$. This vector space robust against collective noise is
called the decoherence free subspace (DFS).

We generate
basis vectors $|S=0, S_z=0 \rangle$ of two
one-dimensional representations of SU(2) from $\{|e_{ai} \rangle,
|e_{bi} \rangle\}$ as
\begin{eqnarray*}
|0 \rangle_L&=&\frac{1}{\sqrt{2}}(|1 \rangle |e_{a1} \rangle - 
|0\rangle \Sigma_x|e_{a1} \rangle)
= \frac{1}{\sqrt{2}}(|1 \rangle |e_{a1} \rangle + 
|0\rangle (\sigma_x)^{\otimes 3}|e_{a1} \rangle),\\
|1 \rangle_L&=&\frac{1}{\sqrt{2}}(|1 \rangle |e_{b1} \rangle - 
|0\rangle \Sigma_x|e_{b1} \rangle)
= \frac{1}{\sqrt{2}}(|1 \rangle |e_{b1} \rangle + 
|0\rangle (\sigma_x)^{\otimes 3}|e_{b1} \rangle),
\end{eqnarray*}
where $\Sigma_x = \sum_{i=1}^3 \sigma_x^i=-\sigma_x^{\otimes 3}$ for $S=1/2$.
It is important in the implementation of the encoding circuit to realize
that
\begin{eqnarray*}
|0\rangle_L &=& (X\otimes I_8)({\rm{CNNN}})(H\otimes I_8)|0 \rangle \otimes
|e_{a1} \rangle = (X\otimes I_8)({\rm{CNNN}})(H\otimes U_E^{(3)})|0 \rangle \otimes
|000 \rangle, \\
|1\rangle_L &=& (X\otimes I_8)({\rm{CNNN}})(H\otimes I_8)|0 \rangle \otimes
|e_{b1} \rangle = (X\otimes I_8)({\rm{CNNN}})(H\otimes U_E^{(3)})|0 \rangle \otimes
|001 \rangle,
\end{eqnarray*}
where CNNN is a controlled NOT gate with one control bit (the top-most qubit)
and three target bits (the rest of the qubits). 

\begin{figure}
\includegraphics[width=8cm]{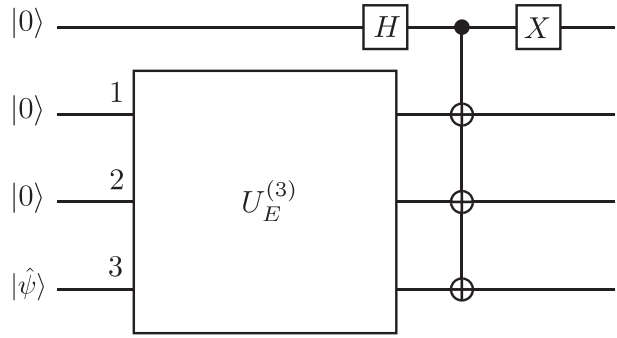}
\caption{Encoding circuit $U_E^{(4)}$
of the decoherence free subspace for a 4-qubit
system. It encodes a single qubit state $|\hat{\psi}\rangle$.
The recovery operation is given by $U_E^{(4)\dagger}$.}\label{fig:4}
\end{figure}

Figure \ref{fig:4} shows an example of the encoding circuit for
the four-qubit DFS. In contrast with the three-qubit noiseless subsystem,
the second qubit (the first input qubit of $U_E^{(3)}$) must be initially set to 
$|0 \rangle$ for successful encoding of the DFS in the present case.

\section{5-qubit Noiseless Subsystem}

Noiseless subsystem using five qubits encodes two data qubits.
It is recursively implemented by employing the encoding
circuit $U_E^{(3)}$ for the three-qubit noiseless subsystem.

The algebra $\mathcal{A}_5$ obtained from ${\bf{2}}^{\otimes 5}$
is decomposed into the sum of irreducible representations as 
$$
{\bf{2}}^{\otimes 5} ={ \bf{6}} \oplus (I_4 \otimes {\bf{4}})
\oplus (I_5 \otimes {\bf{2}}).
$$
Corresponding to this decomposition, any unitary matrix 
$V \in {\bf{2}}^{\otimes 5}$ is decomposed as
$$
V = V_6 \oplus (I_4 \otimes V_4) \oplus (I_5 \otimes V_2)
$$
under a proper choice of basis vectors. Here $V_k$ belongs to the
irreducible representation $\bf{k}$,
$k=2, 4, 6$ of SU(2). We implement a noiseless subsystem 
by employing the five two-dimensional representation spaces.

Let $\{|e_{ai} \rangle, |e_{bi} \rangle\}$ be basis vectors
introduced in Section II. We generate
four basis vectors $\{|00 \rangle_L, |01 \rangle_L,|10 \rangle_L,
|11 \rangle_L\}$ from four
two-dimensional representations of SU(2) as
\begin{eqnarray*}
|00 \rangle_L&=&\frac{1}{\sqrt{2}}(
|01 \rangle -  |10\rangle ) |e_{a1} \rangle,\\
|01 \rangle_L&=&\frac{1}{\sqrt{2}}(|01 \rangle
- |10\rangle) |e_{b1} \rangle,\\
|10 \rangle_L &=&  \frac{1}{\sqrt{6}}(|01 \rangle
+ |10 \rangle)|e_{a1} \rangle -2 |00\rangle
|e_{a2} \rangle,\\
|11\rangle_L &=&  \frac{1}{\sqrt{6}}(|01 \rangle
+ |10 \rangle)|e_{b1} \rangle -2 |00\rangle
|e_{b2} \rangle.
\end{eqnarray*}
It is important to realize the self-similar structure between the
above basis vectors and those of the 3-qubit noiseless subsystem.
The third qubit basis vectors in the latter case is replaced by the 
logical qubit basis vectors of the 3-qubit noiseless subsystem in
the above basis vectors. This observation makes implementation
of the encoding/decoding circuit almost a trivial work.
Note that we do not need to worry about the rest of the basis
vectors so far as they are orthogonal to the above basis vectors
spanning the noiseless subsystem and that this orthogonalization
is automatically taken into account if we employ the unitary matrix 
$U_E^{(3)}$ for implementation.

\begin{figure}
\includegraphics[width=9cm]{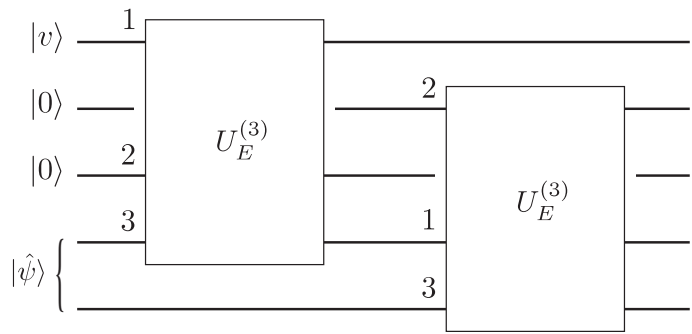}
\caption{Encoding circuit $U_E^{(5)}$ of the 5-qubit noiseless
subsystem, which encodes a two data qubit state
$|\hat{\psi}\rangle$.}\label{fig:5}
\end{figure}

Figure \ref{fig:5} shows an example of the encoding circuit of 
the five-qubit noiseless subsystem. It should be noted that 
the top-most qubit can be any state while all the other encoding
ancilla qubits must be in the state $|0 \rangle$. Each $U_E^{(3)}$
acts on the three qubits numbered 1, 2 and 3, which are fed into
the input ports 1,2 and 3, respectively, in Fig. 1.
The qubit line passing underneath the gate $U_E^{(3)}$ is
not affected by $U_E^{(3)}$. 

\section{Summary and Discussions}

Decoherence free subspace (DFS) and noiseless subsystem make use of
vector subspaces which are
immune to collective noise of the form $W^{\otimes n}$,
where $W$ belongs to $\bf{2}$ of SU(2). We have constructed
simple encoding and decoding quantum circuits of noiseless
subsystem for $n=3$ and $5$ and DFS for
$n=4$. Our strategy is to use the encoding/decoding
circuit $U_E^{(3)}$ for $n=3$ recursively in the implementation
for $n=4$ and $n=5$. 

It can be shown generally
that $m$ logical qubits
are implemented with $(2m+1)$-qubit and 
$(2m+2)$-qubit
systems by the same recursive implementations. 
It should be clear form our construction that $m$ logical
qubits are implemented by use of $m$ $U_E^{(3)}$-modules,
which shows that the circuit complexity for our encoding
and decoding circuits increases merely linearly in $m$. 

Note, however, that our construction is not the most
economical one. There are $\binom{n}{m}-
\binom{n}{m-1}$ basis vectors in 2-dimensional irreducible representations
for $n=2 m+1$, which encode $k = \lfloor \log_2 
\left(\binom{n}{m}-\binom{n}{m-1}\right)\rfloor$ qubits. 
This number $k$ is greater than $m$ for $n \ge 9$, and actually
$k/n \to 1$ as $n \to \infty$. This asymptotic behavior is
also observed in \cite{Kempe} for DFS.

It was shown that the top-most qubit in Figs. \ref{fig:3} and
\ref{fig:5} can be any state.
Although the entropy of the qubit system increases in
general, it remains constant if the top-most qubit
is maximally mixed initially as
$\rho_a= \frac{1}{2} I_2$.
This state is attained after operations of many random unitary errors 
$W^{\otimes n}$, for example. This behavior is somewhat analogous to DFS 
with $\rho_a=|0 \>\< 0|$, in 
which the entropy does not change at all.

\section*{Acknowledgement}

CKL was supported by a USA NSF grant, a HK RGC grant,
the 2011 Fulbright Fellowship, and the 2011 Shanxi 100 Talent Program.
He is an honorary professor of University of Hong Kong,
Taiyuan University of Technology, and Shanghai University.
MN and HT were supported by ``Open Research Center'' Project
for Private Universities: matching fund subsidy from MEXT
(Ministry of Education, Culture, Sports, Science and Technology).
MN would like to thank partial supports of Grants-in-Aid for Scientific 
Research from the JSPS (Grant No.~23540470).
YTP was supported by a USA NSF grant.
NSS was supported by a HK RGC grant.

\end{document}